\documentclass[3p, review,12pt]{elsarticle}

\usepackage[colorinlistoftodos]{todonotes}
\usepackage{placeins}
\usepackage{graphicx}
\usepackage{epsfig}
\usepackage{ecrc}

\usepackage{tikz}
\usetikzlibrary{shapes}
    \newcommand{\markergraycircle}{\raisebox{0.5pt}{\tikz{\node[draw,scale=0.4,circle,fill=gray!100!](){};}}}
    
    \newcommand{\markertriangleunfilled}{\raisebox{0pt}{\tikz{\node[draw,scale=0.3,regular polygon, regular polygon sides=3,fill=none](){};}}}
    \newcommand{\markerdiamondunfilled}{\raisebox{0pt}{\tikz{\node[draw,scale=0.4,diamond,fill=none](){};}}}
    
    \newcommand{\markersquareunfilled}{\raisebox{0.5pt}{\tikz{\node[draw,scale=0.4,regular polygon, regular polygon sides=4,fill=none](){};}}}
    
\volume{00}
\firstpage{1}
\journalname{Journal of Membrane Science}
\runauth{D. Wypysek}
\jid{J. Memb. Sci.}
\jnltitlelogo{J. Memb. Sci.}

\usepackage{siunitx} 
\usepackage{verbatim} 
\usepackage{subfig} 
\usepackage{microtype} 

\usepackage{amssymb}
\usepackage[version=4]{mhchem}
\usepackage{overpic}
\usepackage{graphics}
\usepackage{multirow}
\usepackage{rotating}
\usepackage{textcomp}
\usepackage{float}
\usepackage{color}

\begin{document}

\begin{frontmatter}

\title{In-situ investigation of wetting patterns in polymeric multibore membranes via magnetic resonance imaging}

\author[DWI,CVT]{Denis Wypysek\corref{authors}}
\author[DWI,CVT]{Anna Maria Kalde\corref{authors}}
\author[CVT]{Florian Pradellok}
\author[DWI,CVT]{Matthias Wessling\corref{mycorrespondingauthor}}

\cortext[authors]{These authors contributed equally to this work.}
\cortext[mycorrespondingauthor]{Corresponding author: manuscripts.cvt@avt.rwth-aachen.de}

\address[DWI]{DWI-Leibniz - Institute for Interactive Materials, Forckenbeckstrasse 50, 52074 Aachen, Germany}
\address[CVT]{RWTH Aachen University, Chemical Process Engineering, Forckenbeckstrasse 51, 52074 Aachen, Germany}

\begin{abstract}
Wetting of the membrane to displace air or conditioning liquids is important to exploit the complex porosity of a filtration membrane. This study reveals the details of wetting in multibore membrane-based filtration modules. Using magnetic resonance imaging (MRI), we quantify the fluid distribution patterns during initial membrane wetting in dead-end permeation mode. The spatio-temporal evolution of aqueous copper sulfate solution wetting the membrane fibers was investigated as a function of the applied flux, packing density, and position along the membrane module length. 

Three initial wetting conditions were examined: delivery-state membranes, ethanol-washed and dried (air-filled) membranes, and ethanol-filled membranes. Significant changes in wetting patterns were observed due to interfacial and polymer swelling effects.
This in-situ investigation reveals a slow wetting progression over six hours and more to obtain complete wetting, even at high fluxes of 200~LMH. However, an increased flux leads to faster wetting kinetics as the evolving wetting patterns are flux dependent.
The packing density of the multibore fibers additionally impacts the wetting kinetics by shifting the prevalent pressure conditions. Although in dead-end mode, the wetting progression is non-uniform along the membrane module length.

In addition to this parameter study, different pre-wetting agents' effect on the displacement behavior was investigated in depth.  
This study helps to understand (a) complex wetting phenomena inside multibore membranes in dead-end filtration, (b) the membranes' interaction with their surroundings due to neighboring membranes, and (c) the effect of the used fluid system for displacement on the resulting wetting patterns. 
\end{abstract}


\end{frontmatter}

\section{Introduction}

    Microfiltration and ultrafiltration membranes possess a porous nature that allows them to separate particular matter from a liquid system. One type of microfiltration membrane is multibore membranes. Multibore membranes are monolithic and house several bore channels within one polymer matrix. The monolithic geometry results in higher mechanical robustness compared to single bore fibers~\cite{Gille.2005, Wang.2014, Wan.2017}. Those improved tensile properties allow multibore membranes for industrial-scale applications in drinking water production and seawater desalination~\cite{Gille.2005, Heijnen.2012, BuRashid.2007}, as well as wastewater and surface water filtration~\cite{Wang.2014, Wang.2012, Teoh.2011, Wang.2014b}.
    
    Multibore membranes have been subject to many studies regarding optimized geometry~\cite{Heijnen.2012, Wang.2014, Teoh.2011, Peng.2011, Luelf.2018b}, membrane material~\cite{Lu.2016, Luo.2014, Luo.2015}, and experimental flow distribution~\cite{Heibel.2003,Zhu.2017, Wypysek.2019b, Schuhmann.2019b,Simkins.2020}, as well as flow distribution via computational fluid dynamics (CFD)~\cite{Wypysek.2019b, Frederic.2018, Kagramanov.2001}. Especially the flow distribution across a multibore membrane has been proven to be very sensitive to non-idealities such as bending, skin damages, or excentric positioning. In turn, this altered flow distribution causes distinct changes in flow patterns across the whole module~\cite{Wypysek.2019b, Schuhmann.2019b}.

    A versatile method for the analysis of fluid flow distribution and the online and in-situ characterization of membrane filtration is magnetic resonance imaging (MRI) and flow-MRI. MRI is a non-invasive tool to monitor fluid velocities and particle depositions. Studies using MRI and flow-MRI for characterization of membrane filtration with liquid-liquid \cite{S.Yao.1995,J.M.Pope.1996} as well as liquid-solid systems \cite{D.Airey.1998,Buetehorn.2011b, Culfaz.2011b} have been presented in literature. Thereby, there is no need for large superconducting magnet MRI systems for the measurement of membrane modules to gain valuable information, as low-field benchtop hardware can be used for this purpose~\cite{YANG.2014, Wiese.2018b}.

    As the MRI technology also allows the measurement of liquid content present in the membrane, wetted and non-wetted areas inside the porous structure can be distinguished. Incomplete wetting of membranes can influence the permeation and performance of the filtration process significantly~\cite{Pena.1993,WANG.2005,Bohonak2005}. Hence, quantifying the degree of wetting and wetting times are of major importance to judge the extent of permeable porosity. The majority of studies must assume that permeation experiments access a completely solvent-wetted membrane porosity. Comprehensive studies presented in the literature perform washing and wetting experiments prior to the final transport studies without further investigation of the achieved wetting degree. Hence, all studies presented in the literature focus on membrane performance in steady-state, or at least on the performance after several usage cycles. However, this initial sub-optimal membrane performance might significantly impact later non-uniform flow distributions and fouling events. As early sub-optimal membrane performance has yet hardly been evaluated, investigating this loss in membrane performance at an early stage of usage is the main aim of the study presented here.
    
    Here, we report MRI inspection studies on the degree of wetting and correlate these with permeation experiments. We investigate the wetting behavior of commercial polyethersulfone (PES) multibore membranes (SevenBore\texttrademark{}) inside custom made membrane modules. The effect of packing density inside the module on the membrane wetting is analyzed in detail. Direct wetting of membranes in delivery conditions is compared to pre-wetted membranes. Wetting degrees and wetting times are measured. With this study, we aim to provide a more detailed understanding of initial permeation conditions and suggest process conditions allowing a full wetting of the membranes, which in turn leads to overall homogenized transport properties. By analyzing wetting and flow patterns at first membrane usage, we strive to understand and quantify the mechanisms causing sub-optimal membrane usage.

\section{Background on wetting}
    Membrane wetting is mostly addressed for membrane contactors~\cite{Ibrahim.2018}, membrane distillation~\cite{Pena.1993, Rezaei.2018, MinweiYao.2020}, microsieves~\cite{Girones.2005}, and monoliths~\cite{Marinkovic2020}. However, literature states that microfiltration and ultrafiltration membrane performance is significantly influenced by its wetting state~\cite{Kochan.2009, Roudman.2000, Argyle.2015, Argyle.2017}. Conventionally, commercially available ultrafiltration membranes are pre-filled with a pore stabilizer such as glycerol~\cite{He.2010}. This fluid needs to be replaced by the feed solution before the filled pore can contribute to the filtration performance.
    
    The results presented in the literature indicate that even at advanced stages of filtration, the membranes remain partially non-wetted with water, especially for hydrophobic membranes in microporous parts of the structure~\cite{Kochan.2009, Argyle.2015}. This incomplete wetting is caused by the non-uniform pore size distribution in polymer networks produced via phase separation. Non-uniform pores that are connected and simultaneously wetted by a fluid form communicating networks. This network communication leads to non-uniform wetting~\cite{Das.2005} and effects like fluid droplet trapping and bypassing flows~\cite{Pinder.2008}. In those cases, the necessary hydrodynamic pressure to wet the pores cannot be reached with the macroscopically applied transmembrane pressure (TMP). Hence, macroscopically non-uniformly wetted structures arise.
    
    On a pore-scale level, the necessary pressure for pore wetting $\Delta p_{B}$ depends on the surface tension $\gamma$, the macroscopic contact angle $\theta$, and the local pore diameter $d_{Pore}$ of the specific membrane and liquid system. It can be derived from the Young-Laplace equation (see Equation~\ref{eqbreakthrough}). The Young-Laplace equation is valid for large pore diameters, as additional effects such as van der Waals and electrostatic double-layer forces can have a dominating effect on the fluid-membrane interface for small pores of only some nanometers~\cite{Butt.2003}. The membranes investigated are ultrafiltration membranes with a nominal pore size of 40~nm. Equation~\ref{eqbreakthrough} is valid for the system used in this study.
    
        \begin{equation}\label{eqbreakthrough}
            \Delta p_{B}=\frac{4\cdot\gamma\cdot cos\theta}{d_{Pore}}
        \end{equation}
    
    At the membrane surface - namely where the retention of particular matter takes place - liquid only penetrates the membrane through wetted pores. Hence, the total liquid transport can be calculated by the sum of all flows through single pores. The volume flow $\dot{V}$ through a single pore can be calculated using the law of Hagen-Poisseuille as described in Equation~\ref{eqHagenPoisseuille}. Here, the flow depends on the pore radius $r$, the local pressure gradient $\Delta p$, the liquid viscosity $\eta$, and the pore length $L$.
     
        \begin{equation}\label{eqHagenPoisseuille}
            \dot{V}=\frac{\pi r^{4}}{8\cdot\eta}\frac{\Delta p}{L}
        \end{equation}
    
    A non-uniformly wetted membrane decreases the flux across the membrane in two ways. First, the number of pores contributing to the flux is reduced. Thus, the overall flux linearly decreases with an increasing number of non-water-wetted pores. Second, a non-uniformly wetted membrane only allows distinct flow paths for the liquid to pass through the membrane. Hence, flow patterns inside the membrane change, which causes unfavorable pathways limiting the liquid mass transport by increased pressure drop.~\cite{Wypysek.2019b}

\section{Materials and methods}

    \subsection{Experimental setup}\label{exSetup}
\begin{figure}[H]
                \centering
                \includegraphics[width=\textwidth]{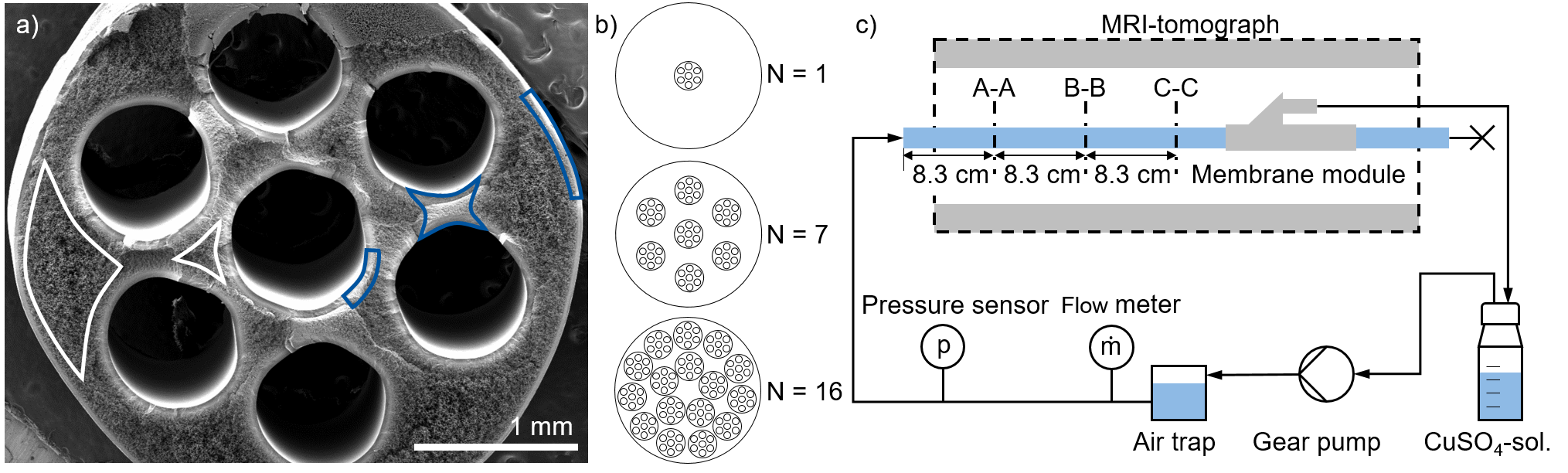}
                \caption{Presentation of (a) a FESEM image of the cross-section of the investigated SevenBore\texttrademark{}\texttrademark{} membrane fibers, (b) three considered membrane packing densities and the correlating membrane arrangements in the membrane module, (c) the experimental setup for the investigation on the wetting behavior of SevenBore\texttrademark{} membranes via MRI.}
                \label{figSetup}
            \end{figure}

        \paragraph{\textbf{Membrane}}
        The investigated inside-out ultrafiltration membranes with a nominal pore size of \SI{40}{\nano\meter} are commercially available PES SevenBore\texttrademark{}~fibers supplied by SUEZ Water Technologies \& Solutions. The membrane cross-section is provided with a central bore, surrounded by six symmetrically arranged bores close to the outer fiber diameter, which is approximately \SI{4}{\milli\meter}. Each bore has a diameter of approximately \SI{0.9}{\milli\meter}. To reveal the membrane pore structure, we used a field-emission scanning electron microscope (FeSEM, Hitachi High-Technologies Corporation Model S-4800) to image the membrane cross-section. Figure~\ref{figSetup}~a) depicts the unique pore structure of the investigated membrane fibers. While blue framed areas mark the active membrane layer with smaller pore sizes, white framed areas depict the areas with bigger pore sizes representing the membrane's support structure. Domains with bigger pore sizes provide mechanical stability of the membrane fiber and do not contribute to particular matter separation. 
        
        \paragraph{\textbf{Module}}
        For the examination of wetting morphologies inside the pore structure, individual fibers must be incorporated into modules. Thereby, three packing densities are investigated, which contain either one, seven, or 16 membrane fibers accounting for packing densities of 5~\%, 35~\%, and 79~\%, respectively. The corresponding membrane arrangements are presented in Figure~\ref{figSetup}~b). We used 3D-printed spacers to ensure constant positioning of the membrane fibers inside the single and seven fiber modules. The manufactured modules are \SI{50}{\centi\meter} in length and have an internal diameter of \SI{18}{\milli\meter}. 
        
        Furthermore, the membrane modules contain one connection for the feed stream and one connection for the permeate and retentate, respectively. The wetting behavior is monitored at three positions along the module length (see Figure~\ref{figSetup}~c)). Distances from the module inlet are \SI{8.3}{\centi\meter} for position A-A, \SI{16.6}{\centi\meter} for position B-B and \SI{25}{\centi\meter} for position C-C. 

        \paragraph{\textbf{Magnetic resonance imaging setup}}
        Figure~\ref{figSetup}~c) represents the developed experimental setup. All MRI measurements are carried out on a Magritek low-field NMR tomography system combined with a KEA\textsuperscript{2} low-frequency spectrometer. To control the hardware, Magritek's software Prospa is used. The tomograph operates with a field strength of \SI{0.56}{\tesla} at a Larmor frequency of \SI{23.8}{\mega\hertz}. Depending on the experimental procedure, MRI-parameters have to be adjusted. Within the \SI{60}{\milli\meter} inner diameter imaging probe, the membrane module is placed horizontally. Measurements in the tomograph are carried out in the x-y-plane, being the cross-section of the membrane module. All images are obtained using a 2D spin-echo pulse sequence adapted from our previous studies~\cite{Wiese.2018b,Wiese.2019b,Wypysek.2019b}. In all experiments, a fluid circuit is established with an ISMATEC\textsuperscript{\textregistered{}} ISM405A gear pump. Thereby, fluxes of 50~liter per square meter membrane area and hour (LMH), 100~LMH, and 200~LMH are applied to the membrane modules. For wetting experiments, a copper sulfate solution (\SI{3.12}{\gram\per\liter} copper(II) sulfate pentahydrate (\ce{CuSO4.5H2O}, $\geq$~99.5\%, Carl Roth, Germany) in de-ionized water) was used as aqueous phase, hereafter termed aqueous solution. This concentration is equivalent to 0.05~M ionic strength. The use of \ce{CuSO4}-solution has a potential influence on the wetting progress as it might shift the contact angle of the used material system. However, \ce{CuSO4} is indispensable for MRI since the required time to capture a magnetic resonance image is significantly reduced. Corresponding mass flow rates across the membrane are monitored with a mini CORI-FLOW\texttrademark{} M14 (Bronkhorst High-Tech B.V., Netherlands). The resulting transmembrane pressure (TMP) is recorded with a Wika pressure transmitter S-11 (Wika, Germany). To ensure comparability between experiments on the impact of flux and packing density, we fabricated an individual module with new membranes for each experiment.

    \subsection{Wetting procedures}\label{WettingProcedures}
        
        Typically, ultrafiltration membranes are factory-filled with a pore stabilizer, as the pore network in the selective layer tends to collapse when dried. According to the manufacturer, glycerol is used as a pre-wetting agent which is known to be a prominent choice for pore network stabilization~\cite{SUEZWaterTechnologies&Solutions.}. Glycerol as a pre-wetting agent can be removed by washing with ethanol. To investigate the amount of contained pre-wetting agent, as well as to evaluate the approach of ethanol-washing, we immersed single membrane fiber samples in \ce{EtOH} for two hours and dried these in an exsiccator for 30~min. The sample masses were compared to factory-filled membranes, also dried in an exsiccator for the same time. The weight loss of 5~cm membrane samples can be seen in Table~\ref{tabweightloss} and confirms that the pre-filling agent is washed-out successfully.  
        
            \begin{table}[H]
                \centering
                \caption{Weight of 5~cm membrane samples for factory-filled membranes, dried membranes, and washed and dried membranes and the corresponding weight loss after treatment.}
                \begin{tabular}{llll}
                \hline
                    Configuration & Mass [mg/cm] & Weight loss [mg/cm]&Weight loss [\%]\\\hline
                    Factory-filled & 35.54 $\pm$ 1.7 & - & - \\
                    Dried & 33.04 & 2.50 & 7.03 \\
                    Washed and dried & 16.94 & 18.60 & 52.34 \\\hline
                \end{tabular}
                \label{tabweightloss}
            \end{table}       
        
        To correlate the MRI data obtained from membrane investigation to the observed changes in mass due to drying, we used MRI reference measurements. The signal strengths of pure fluids, as well as different fluid systems inside a membrane structure, were compared to show the suitability of MRI measurements to distinguish wetting states inside the membrane. Details of those reference experiments are shown in Supplement Section~2. In fact, this data implies that main parts of the membrane are factory-filled with a pore-stabilizing agent that can be removed with ethanol-washing. We, hence, conclude that at least the selective layer of the investigated SevenBore\texttrademark{} membranes is fully pre-filled with a pore-stabilizing agent. However, no definite conclusion can be drawn about the wetting state of the more macroporous support structure. Based on the findings of sample masses and MRI references, three sets of experiments were carried out to investigate the impact of the wetting fluid on the wetting morphologies inside the pore structure via MRI.
        
        First, the pre-wetting agent in the pore structure of commercially available membrane fibers is displaced with the aqueous solution. We assume the pre-wetting agent to be present at least in all pores of the selective layer, and to some extent in the mechanical support domains. 
        
        Second, the pre-wetting and stabilizing agent is removed by ethanol washing. Throughout ethanol washing, seven membrane fibers are immersed in approximately \SI{150}{\milli\liter} ethanol (\ce{EtOH}, $\geq$ 96~\%, Carl Roth, Germany) for 24~hours. Afterward, the washed fibers are dried by placing them in a fume hood for at least 24~hours at room temperature and ambient pressure. Washed and dried membrane fibers are incorporated into modules containing seven fibers and are wetted with the aqueous solution.  
        
        Third, membranes are washed with ethanol directly inside the module to replace the prevalent glycerol in the pores entirely. Without any drying step, the ethanol is displaced by aqueous solution. For ethanol wetting, an ethanol flux of 100~LMH is applied to a membrane module containing seven SevenBore\texttrademark{} fibers. Thereby, the progress of ethanol wetting is monitored via MRI. With the fibers thoroughly wetted with ethanol, the fluid is switched to aqueous solution at a flux of either 50~LMH or 100~LMH. 
        Each wetting experiment was conducted in dead-end mode for approximately 6~hours. Depending on the material system, the MRI-parameters vary to obtain a maximum spatial and temporal resolution. Thus, imaging times varied from 4:47~min for factory-filled, as well as for washed and dried membranes to 6:50~min for ethanol-filled membranes leading to time-averaged magnetic resonance images. MRI-parameters for all three material systems are listed in Table~\ref{tabparameters}.
        
            \begin{table}[H]
                \centering
                \caption{MRI-parameters for the investigation of wetting morphologies in factory-filled membrane fibers, washed and dried fibers, and ethanol-filled fibers with \ce{CuSO4}-solution.}
                \begin{tabular}{lllll}
                \hline
                    Wetting fluid in pores & & Glycerol & Air (washed \& dried) & Ethanol\\
                    Displacing fluid & & \ce{CuSO4}-sol. & \ce{CuSO4}-sol. & \ce{CuSO4}-sol.\\\hline
                    Parameter & Unit & Value & Value & Value \\\hline
                    Repetition time & [ms] & 280 & 280 & 400 \\
                    Echo time & [ms] & 48 & 48 & 32 \\
                    Field of view Read [x] & [mm] & 24 & 24 & 45 \\
                    Field of view Phase [y] & [mm] & 24 & 24 & 40 \\
                    Slice thickness [z] & [mm] & 10 & 10 & 10 \\
                    Number of pixels [x\,$\times$\,y] & [\,-\,] & 256\,$\times$\,256 & 256\,$\times$\,256 & 256\,$\times$\,256 \\
                    Resolution [x\,$\times$\,y] & [µm/pixel] & 94\,$\times$\,94 & 94\,$\times$\,94 & 176\,$\times$\,156 \\
                    Number of scans & [\,-\,] & 4 & 4 & 4 \\
                    Acquisition time & [min:sec] & 4:47 & 4:47 & 6:50 \\\hline
                \end{tabular}
                \label{tabparameters}
            \end{table}

     \subsection{Data processing for wetting quantification} \label{WettingDegree}

        The obtained MRI raw data consist of 512x512 pixel images (due to zero filling) that depict a normalized signal strength. Thereby, the measured signal strength of a voxel is assigned to the corresponding pixel. The highest signal is obtained by water. The higher the water content inside a measured voxel, the higher the signal strength. Other hydrogen-containing substances, such as glycerol and ethanol, produce a lower signal. No signal, depicted in black pixels, is obtained when no hydrogen bonds can cause a signal in the corresponding voxel.
        
\begin{figure}[H]
            \centering
            \includegraphics[width=0.9\textwidth]{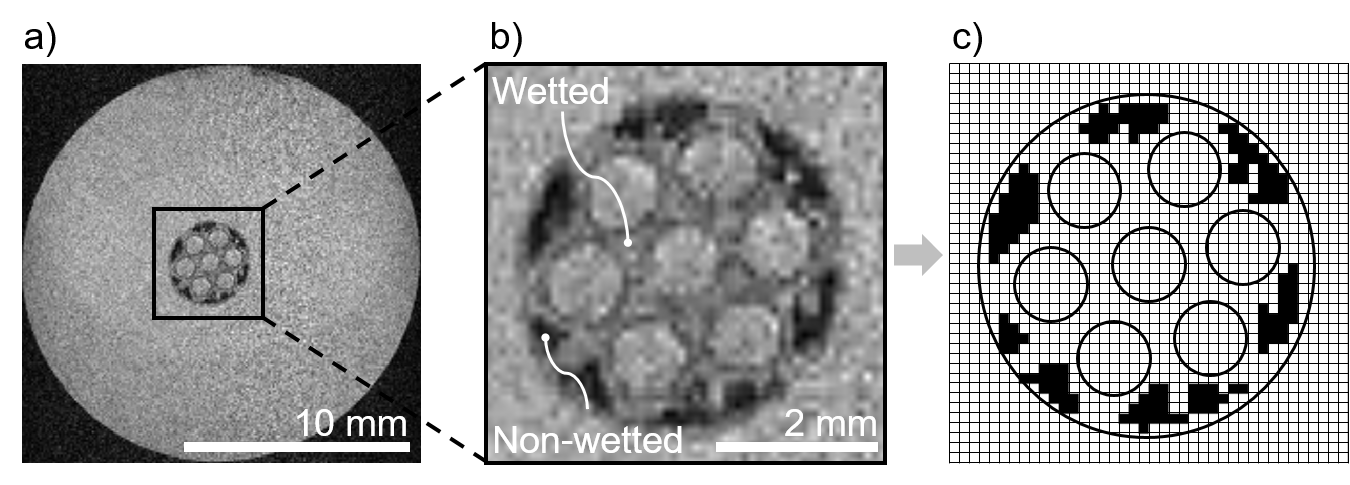}
            \caption{Representation of the functionality of the developed MatLab tool for the determination of the membrane's wetting degree. a) Magnetic resonance image of a single fiber module during the wetting process. b) Magnified image of the membrane fiber with wetted (greyish) and non-wetted (blackish) areas in the membrane cross-section. c) Discretization performed by the developed MatLab tool in wetted and non-wetted pixels.}
            \label{figMethodology}
        \end{figure}

        Figure~\ref{figMethodology}~a) shows an exemplary raw imaged during the wetting procedure of a single fiber module. The light outer domain represents the shell side of the module filled with aqueous solution, while the middle darker structure depicts the SevenBore\texttrademark{} membrane. Figure~\ref{figMethodology}~b) shows a magnification of the membrane area. Different values on the grayscale can be distinguished. The seven light circles depict the seven lumen channels of the membrane. However, also inside the polymer structure, different signal strengths are observed. This difference in signal strength results from a combination of aqueous solution wetted (brighter) or non-wetted (darker) sections inside the membrane.
        
        With the obtained raw data, the membranes' wetting degree (WD) can only be compared qualitatively. For quantitative analysis, image processing is needed. Hence, a MatLab code was developed. Here, the membrane area was defined automatically. Within this area, setting a threshold for single pixels' signal strength lead to a binary image. To identify a reasonable threshold, a membrane was fully wetted with aqueous solution, and the MRI signal strength within the wetted membrane matrix was measured. This signal strength was then set as the threshold for the distinction of wetted and. A more detailed sensitivity analysis of this threshold is given in Supplement Table~2. Light pixels were attributed as wetted membrane area, black pixels as non-wetted membrane area. Figure~\ref{figMethodology}~c) shows an exemplary binary image. The wetting degree was determined by the ratio of wetted area to the total membrane area (see equation~\ref{eqWD}). 
        
        \begin{equation}
        \label{eqWD}
           \textrm{Wetting degree WD}
           =\frac{\sum \textrm{Wetted pixels}}{\sum \textrm{Wetted pixels} + \sum \textrm{Non-wetted pixels}}
        \end{equation}

\section{Results and discussion}

    In this study, the wetting progress in SevenBore\texttrademark{} membranes with three different pre-wetting agents is investigated, which are the factory-filled pore stabilizer glycerol, air, and ethanol. We present both qualitative and quantitative analysis. 

    \subsection{Aqueous solution wetting delivery-state membranes}
        \label{secGlycfilled}
        \paragraph{\textbf{Single fiber wetting}}
        
\begin{figure}[H]
                \centering
                \includegraphics[width=0.71\textwidth]{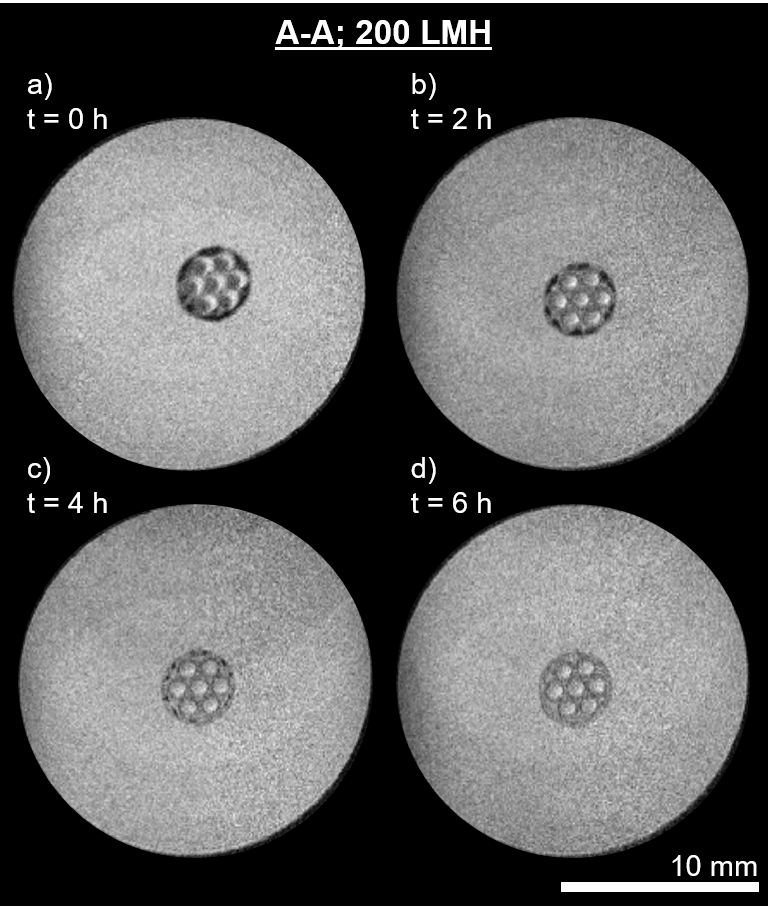}
                \caption{Magnetic resonance images of a delivery-state single fiber module during the wetting process at 200~LMH at position A-A (a) in the beginning of the experiment, (b) after two hours, (c) after four hours, and (d) after six hours of permeation.}
                \label{figSingleFibre}
            \end{figure}

        Figure~\ref{figSingleFibre} shows the progression of membrane wetting over time for a single fiber module at a flux of 200~LMH directed inside-out. The signals were detected close to the feed connection of the module (position A-A).
        
        The image depicted in Figure~\ref{figSingleFibre}~a) is taken directly after filling the module with the aqueous solution and starting the permeation experiment. The membrane area appears to be non-wetted with aqueous solution. The membrane is surrounded by the electrolyte but not positioned centrally. When compared to images taken later during the wetting progress, the membrane is initially found in a different position in the module cross-section. This shift in position is induced by swelling of the polymer material that causes strains and bending of the membrane.~\cite{Wypysek.2019b} The lumen channels cannot be seen as circular light structures but show crescent-shaped light areas. One reason is the membrane's movement that causes artifacts in the MRI measurements and thereby the same position of the light areas inside the lumen channels. 
        
        Throughout the permeation experiment presented in Figure~\ref{figSingleFibre}, a brightening of the membrane area is detected, which expands from the lumen channels outwards. This brightening depicts the wetting areas inside the membrane polymer structure. Thereby, domains associated with smaller pore sizes, as demonstrated in \ref{figSetup}~a), tend to be wetted earlier during the permeation process than domains with larger pore sizes. This is unexpected as the Young-Laplace equation predicts earlier wetting of bigger pore sizes. Additional to the wetting of smaller pores, flow paths through domains associated with larger pore sizes have formed after two hours.
        
        However, even after four hours of permeation, macroscopic non-wetted areas can still be detected, as shown in Figure~\ref{figSingleFibre}~c). Only after approximately six hours, no distinct non-wetted areas remain. This period is in the same range as filtration cycles in commercial applications of SevenBore\texttrademark{} fibers. Hence, the initial effect of membrane wetting might significantly impact the processes' filtration performance and potentially exacerbate fouling and inhomogeneous backwashing phenomena in the industrial-scale applications.
       
        We hypothesize the unexpected chronology of pore wetting to be caused by the used setup control and the series connection of multiple pore sizes along the fluid pathway. In the setup, the applied pressure is adjusted so that a constant flux across the membrane is achieved. Thereby, the pressure distribution inside the membrane fibers varies over the cross-section, as shown in previous studies~\cite{Wypysek.2019b}. High pressure gradients are observed at the membrane interfaces, while a low pressure drop within the support structure is monitored. However, no study has been presented so far on the dynamic pressure distribution during initial membrane wetting. 
        
        To form a wetted pathway from the lumen to the shell side, a series of pores with different pore sizes needs to be wetted subsequently. First, the pores of the active layer at the lumen interface need to be wetted. Second, the larger pores of the support regions need to be traversed. Third, the pores at the shell side interface are flown through. In this last region, narrow pores are predominant due to the phase inversion process used for membrane fabrication. Besides surface energies, viscous forces to displace the higher viscous glycerol need to be overcome. These viscous resistances decline during the displacement process. This is why, initially, a higher pressure needs to be applied to obtain the desired flux. Once the pores of the active layer on the lumen side are wetted with aqueous solution, a lower number of pores in the support domain wetted is necessary to obtain the desired flux as stated by Equation~\ref{eqHagenPoisseuille}. In contrast, numerous pores in the membrane shell need to contribute to fluid transport. Presumably, the pressure distribution during the dynamic membrane wetting also causes steeper gradients at the layers with smaller pore sizes. Hence, a lower driving force for fluid displacement is expected in the support structure, causing the at first unexpected succession of pore wetting. Longer-term wetting of the support domain is then probably driven by incidental interface instabilities and dissolving effects.
        
        \paragraph{\textbf{Packing density}}
        
\begin{figure}[H]
                \centering
                \includegraphics[width=\textwidth]{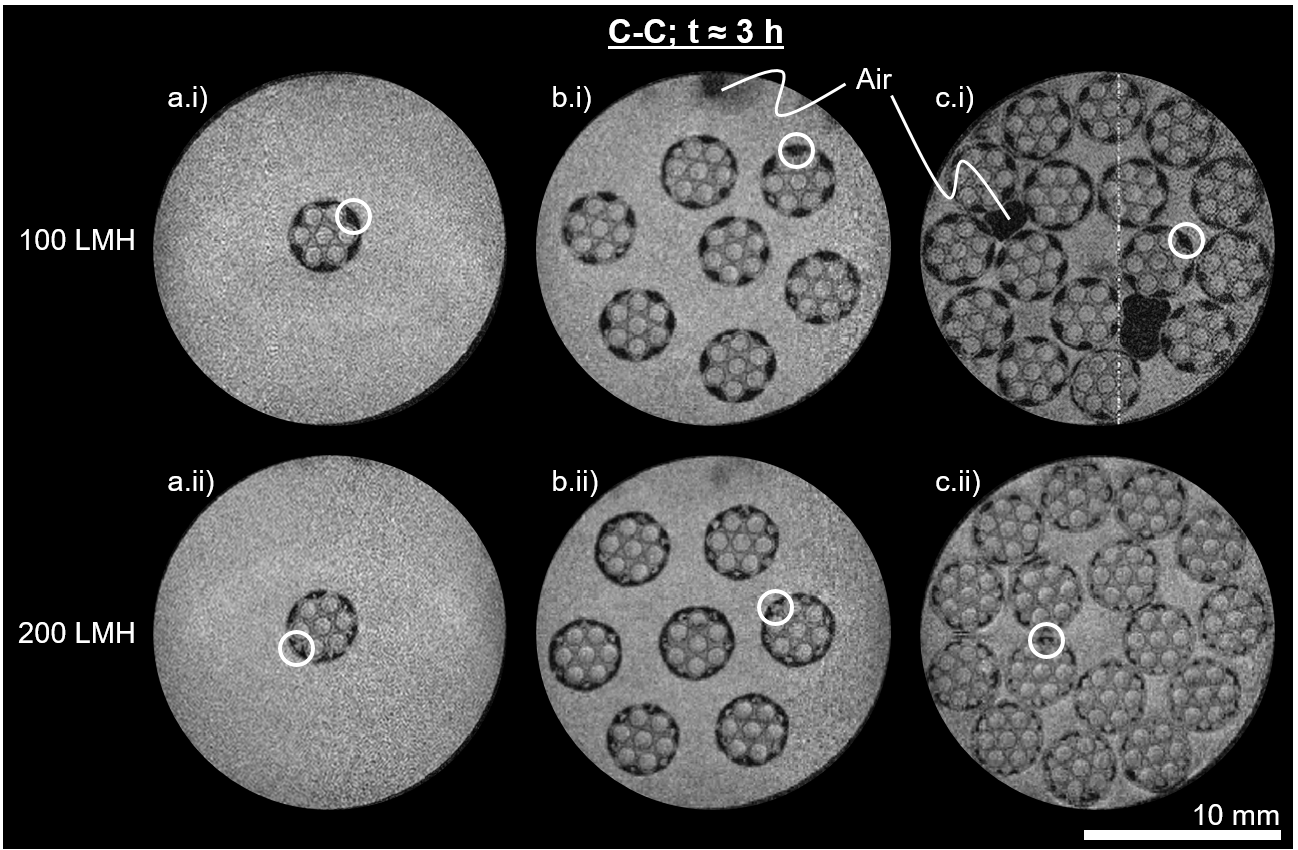}
                \caption{Magnetic resonance images of single fiber modules (a.i-ii), seven fiber modules (b.i-ii) and 16 fiber modules (c.i-ii) containing delivery-state membranes at position C-C taken after approximately 3~hours during the wetting process at 100~LMH (top) and 200~LMH (bottom) with flow rate dependent differences in the wetting behavior highlighted by white-circled areas.}
                \label{figFlowPaths}
            \end{figure}

        Previous studies showed the sensitivity of flow patterns inside membranes and modules towards small pressure disturbances~\cite{Wypysek.2019b, Schuhmann.2019b}. Pressure disturbances increase with non-uniformity in the flow, which can be induced by changing packing density~\cite{Zhang.1995,LAUKEMPEROSTENDORF.1998,Hirano.2012,YANG.2014}. Therefore, we investigated the influence of membrane packing density on the initial membrane wetting behavior. Magnetic resonance images for one, six, and 16 fiber modules containing membranes in delivery-state are displayed in Figure~\ref{figFlowPaths}. Experiments were conducted for 100~LMH and 200~LMH, respectively. The signals were obtained close to the permeate outlet at the C-C plane defined in Figure~\ref{figSetup}~c). As initial wetting state and entirely wetted membranes have the same appearance in MRI for all packing densities, we chose to compare the wetting pathways at their maximum divergence, which is at approximately three hours of permeation. 

        Considering the wetting progression of multiple packing densities at 100~LMH, it can be noted that with increasing packing density, air entrapment between the fibers on the shell side becomes more prominent (no residual air in Figure~\ref{figFlowPaths}~a.i), air at the top of the module in Figure~\ref{figFlowPaths}~b.i), air entrapment between fibers in Figure~\ref{figFlowPaths}~c.i)). This air might accumulate due to incomplete module flooding, or it might be displaced air from the lumen channels, pores that are not wetted with the pre-wetting agent, or temperature enhanced degassing effects, respectively. The latter two phenomena might especially arise if the porous network structure is not fully filled by the pre-wetting agent in delivery state. In this case, most probably air is contained in the larger pores of the mechanical support domains. This might induce additional resistances for displacement and have significant effects on the evolving flow patterns.
        
        However, it can be noted that the wetted pattern inside the membranes looks the same for one, seven, and 16 membranes at 100 LMH. The area surrounding the middle lumen channel in each membrane is fully wetted, whereas the outer regions of the fibers' support structure and a large part of the area close to the outer membrane skin remain non-wetted. The white circles in Figure~\ref{figFlowPaths}~(top) indicate these non-wetted areas. In contrast, the areas where the outer lumen channels are closest to the outer membrane skin are wetted with aqueous solution. Since the aqueous phase can only pass the membrane through wetted areas, we assume that the total feed flux across the membrane is transported through these small wetted parts of the membrane. When comparing this to the results presented in our previous paper \cite{Wypysek.2019b}, and by Schuhmann et al. \cite{Schuhmann.2019b}, a loss in membrane performance as well as an inhomogeneous flow distribution due to incomplete wetting can be assumed, as all parts of the membrane significantly contribute to the overall mass transport when in a fully wetted steady state. 
        
        Applying a flux of 200~LMH to the same number of membranes in the same module geometry has a significant effect on the evolving wetting patterns, as can be seen in Figure~\ref{figFlowPaths}~a.ii), b.ii) and c.ii). In general, a trend of a higher wetting degree with increasing packing density is observed. While the area surrounding the inner lumen channel is fully wetted for all packing densities, especially for the module containing 16 fibers displayed in Figure~\ref{figFlowPaths}~c.ii) the increase in wetting degree is very prominent. This increasing wetting degree can be reasoned with the effect of changed shell side flow conditions in the module and a generally higher transmembrane pressure level for higher packing densities. Although maintaining a constant flow per membrane area, the permeate flow per shell side volume drastically increases. In turn, this leads to higher flow velocities at the shell side. Combined with an increased feed pressure level to supply a sufficiently high mass flow to the lumen channels, this leads to a variation of pressure conditions with changes in the packing density. However, the prevalent pressure conditions mainly determine the local wetting behavior, as described in Equation~\ref{eqbreakthrough}. Hence, the breakthrough pressure is exceeded in more pores for higher packing densities at sufficiently high overall flow rates.
        
        When comparing the wetting patterns of the permeation experiments conducted at 100~LMH and 200~LMH, a distinct difference can be seen for all packing densities. The fibers' support structure's outer regions are wetted when applying a flux of 200~LMH, whereas those domains were non-wetted for a lower flux of 100~LMH. Again, these domains are highlighted by white circles in the bottom row of Figure~\ref{figFlowPaths}. The wetting of these domains majorly contributes to the enhanced wetting degree for the high packing density module at 200~LMH. As described above, an elevated pressure level, also at the pore scale, seems most likely to cause this effect. However, no wetted areas of the fibers' outer skin can be detected for the seven fiber module at 200~LMH flux, whereas there are wetted areas next to the outer lumen channels for a lower flux. One potential reason for this unexpected wetting pattern is that the displacement of the glycerol probably occurs in a radial direction. Therefore, the glycerol formerly trapped in the polymer structure is displaced towards the outer membrane skin. 
        
\begin{figure}[H]
                \centering
                \includegraphics[width=\textwidth]{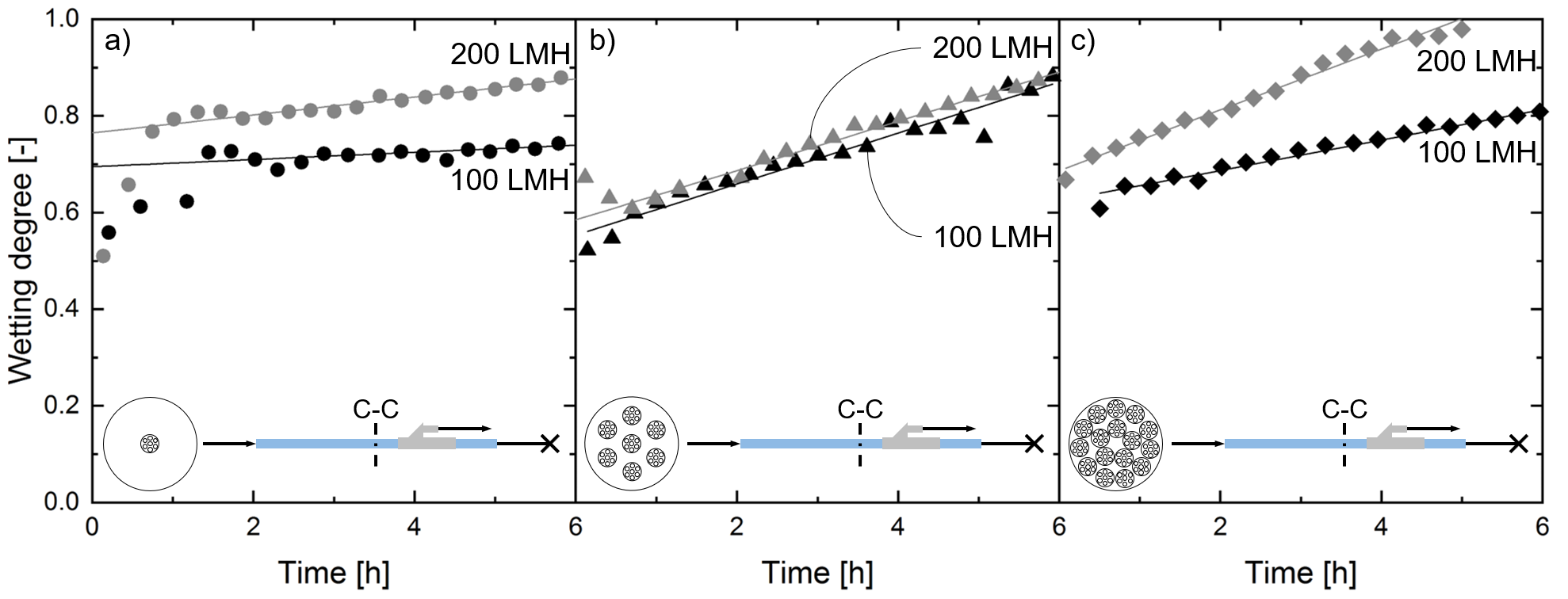}
                \caption{Evaluation of the wetting degree over time at 100~LMH and 200~LMH at position C-C of (a) single fiber modules, (b) seven fiber modules, and (c) 16 fiber modules containing delivery-state membranes including linear curves fitted to the progress of the wetting degree.}
                \label{figWettingPD}
            \end{figure}

        As the data presented in Figure~\ref{figFlowPaths} only displays a temporal snapshot during the wetting process, no conclusion can be drawn from this data regarding the exact evolution of these wetting patterns. To investigate the temporal progress of the membrane fiber wetting, the full data sets of the experiments displayed in Figure~\ref{figFlowPaths} were processed with the methods discussed above. This processed data is displayed in Figure~\ref{figWettingPD}. Permeation experiments were performed over six hours each. The initial wetting degree is higher than zero due to spontaneous wetting during the module's shell-sided flooding and by temporal resolution limitations. Although there is no concrete physical motivation for a linear correlation between wetting degree and time, we chose to indicate linear trend lines to allow an easier comparison of trends. 
        
        Although we present only one experiment per configuration, we assume the observed results to be representative as we see clear trends between several experiments. A high number of manual steps is needed for module assembly, which in itself causes slight variations in the setup. This is why some variations in the observed wetting degrees between single experiments are to be expected. However, the characteristics of the applied hydrodynamic stresses remain the same. Considering different membrane fibers in one module, also these fibers are all subject to similar hydrodynamic stresses. We calculated the wetting degrees of all single fibers in Figure~\ref{figFlowPaths}. Here, all phenomenological wetting patterns observed in the membrane fibers are similar within one experiment. This is coherent with the wetting degrees calculated for the single fibers. These wetting degrees differ by only three to five percent, as shown in Supplement Section~4. Hence, this observation indicates that the formed wetting pathways are reproducible for similar applied hydrodynamic stresses.
        
        For the single fiber (Figure~\ref{figWettingPD}~a)) and the 16 fiber module (Figure~\ref{figWettingPD}~c)), a significantly higher wetting degree throughout the permeation experiments is observed for higher flow rates. However, no such difference can be detected for a seven fiber module (Figure~\ref{figWettingPD}~b)). The wetting degrees calculated for the single fiber module and the seven fiber module show deviating progress at the beginning of the experiments compared to higher packing densities. One potential reason is the initial movement of the membrane due to swelling and bending. This movement causes indistinct MRI signals, which leads to an inaccurate evaluation of the wetting degree. These effects of membrane movement are less pronounced for higher packing densities than for single fibers due to spatial limitations. Nevertheless, the very similar data obtained for seven fiber modules at different fluxes cannot yet be fully explained. 
        
        Figure~\ref{figWettingPD}~c) displays the wetting degree over time for a 16 fiber module. Here, the different slopes of the trend lines for 100~LMH and 200~LMH are prominent. As discussed above, this might be reasoned with the overall higher pressure level used for 200~LMH permeation. In general, it should be noted that the highest possible packing density of 16 fibers in the module at 200~LMH is the only configuration that reached a fully wetted state within the six hours of transport study.

        \paragraph{\textbf{Wetting progression along the module length}}
        
\begin{figure}[H]
                \centering
                \includegraphics[width=\textwidth]{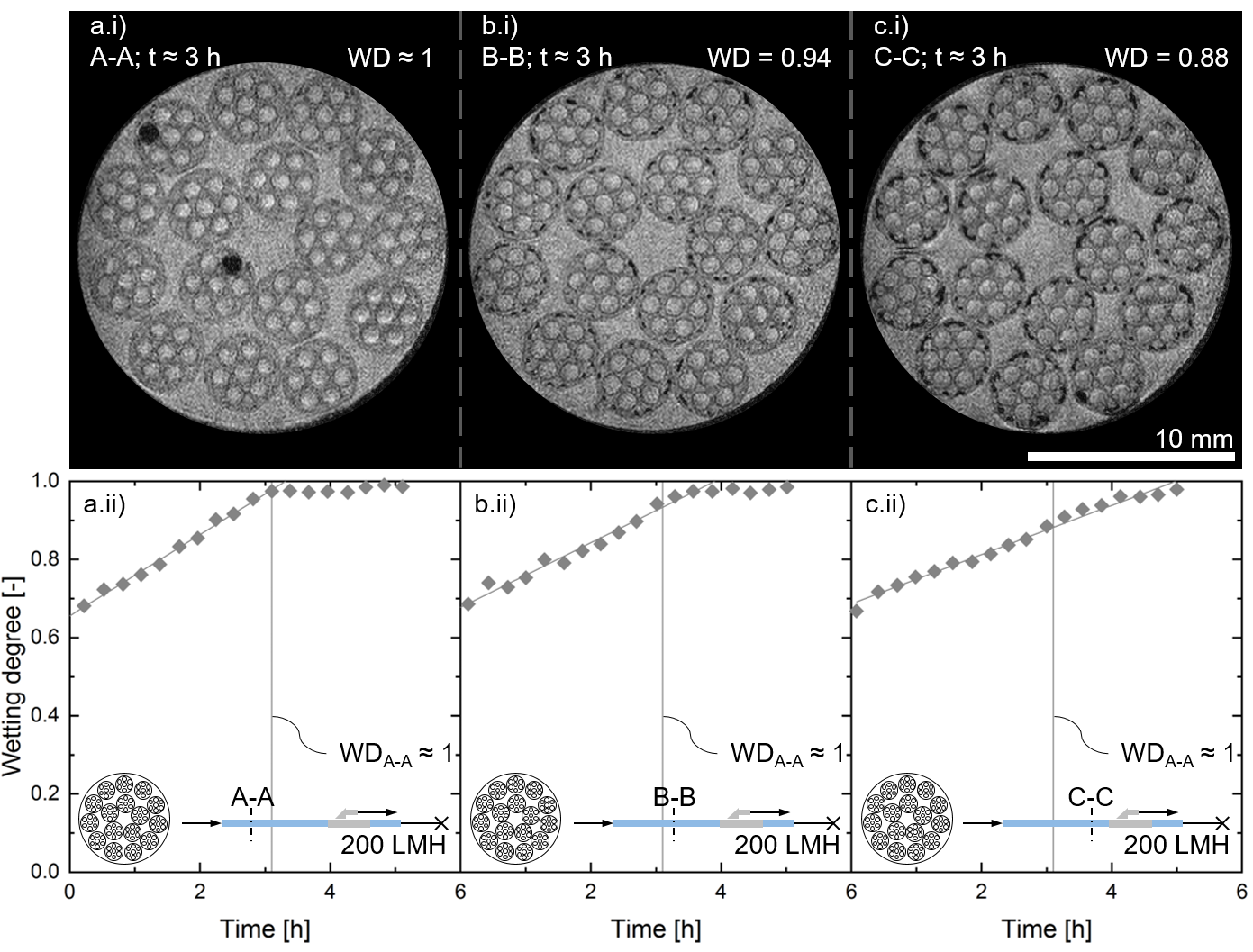}
                \caption{Presentation of (top) magnetic resonance images of a 16 fiber module containing delivery-state membranes as complete wetting was obtained at 200~LMH after approximately 3~hours at position A-A (a.i) and (bottom) the corresponding evaluation of the wetting degree (dots) and the fitted linear function (line) at position A-A (a.ii). No complete wetting was achieved at position B-B (b.i) and C-C (c.i) at this point in time. The slope of the fitted linear functions decreases from position A-A (a.ii) to C-C (c.ii).}
                \label{figWettingPos}
            \end{figure}

        The dominating transport direction inside the porous structure is in a radial direction. However, due to axial flows of feed and permeate streams as well as inhomogeneities in the module like excentric fiber positioning, the flow conditions change along the module length~\cite{Wypysek.2019b}. Therefore, we investigated the wetting state of a 16 fiber module at the three positions A-A, B-B, and C-C as defined in Figure~\ref{figSetup}~c). During one permeation experiment, we alternatingly measured the MRI signals at all three positions, starting at position C-C and moving towards the feed inlet. As one MRI measurement takes approximately five minutes, one data point was taken every 15 minutes at every position. The results of this experiment are displayed in Figure~\ref{figWettingPos}. 
        
        Figures~\ref{figWettingPos}~a.i), b.i), and c.i) show MRI signals of the three positions after approximately three hours of permeation at 200~LMH flux. Wetting degrees at this point in permeation time are indicated in the upper right corner of Figures~\ref{figWettingPos}~a.i), b.i), and c.i). At position A-A, displayed in Figure~\ref{figWettingPos}~a.i), the polymer matrix of the membrane fibers appears fully wetted. Only two lumen channels are filled with air, which can be caused by an unintentional intake of single air bubbles via the feed flow. At position B-B, displayed in Figure~\ref{figWettingPos}~b.i), some domains in the region of the outer membrane skin remained non-wetted. As observed before in Figure~\ref{figFlowPaths}, these areas tend to be between two lumen channels at the outer membrane skin, whereas the outer membrane skin appears to be wetted with aqueous solution at domains in the proximity of lumen channels. At position C-C, displayed in Figure~\ref{figWettingPos}~c.i), the non-wetted domains already seen in Figure~\ref{figWettingPos}~b.i) are more pronounced. Hence, the fiber wetting is less progressed with increasing distance from the feed inlet. 
        
        This trend of decreased wetting degree with increasing distance from the feed inlet can also be seen in the temporal resolution of the observed wetting degree at the three monitored positions in the module, shown in Figures~\ref{figWettingPos}~a.ii), b.ii), and c.ii). At all three positions, the initial wetting degree at the beginning of the permeation experiment is almost identical. As discussed above, the initial wetting degree is significantly higher than zero due to initial wetting during the module flooding by capillary effects and small pressure fluctuations. Also, the effects taking place in the first contact with the liquid system cannot be resolved in MRI measurements due to the measuring time of approximately five minutes and the signal averaging over the full measuring time. 
        
        Linear trends of the wetting degrees over time are indicated. It can be noted that the slope of these trends decreases with increasing distance from the feed inlet. At position A-A, depicted in Figure~\ref{figWettingPos}~a.ii), full wetting is obtained after approximately three hours. For reference, this point in permeation time is indicated with a vertical line in all three graphs showing the wetting degree over time. At positions B-B and C-C, the fibers are not fully wetted after three hours.  While a complete wetting was obtained shortly after at position B-B, a wetting degree close to one at position C-C could only be obtained after more than five hours of permeation.
        
        The difference in wetting kinetics along the module length is puzzling at first sight. Since the experiments are conducted in dead-end mode, little pressure drop along the lumen length is expected. Also, the membrane fibers are expected to be uniform in the module length direction. Additionally, the membrane packing density prevents strong membrane bending and movement inside the module. Therefore, uniform flow distribution and, hence, uniform wetting is expected over the module length. However, two phenomena can be hypothesized that lead to this wetting gradient over the module length. 
        
        First, pressure drops occur both at the shell and the lumen side. We expect the pressure drop inside the lumen channels to be higher due to their smaller flow diameter and, more importantly, the used dead-end permeation mode. This means that the applied TMP as the driving force for membrane wetting is non-uniform over the membrane length, which is well in line with the observation of a decrease in lumen side velocity over the filtration length of the membrane fiber in dead-end mode~\cite{Wypysek.2019b}. This would explain an earlier wetting at the inlet of the membrane module. We hypothesize that the axial wetting progression then is caused by stochastic events that locally exceed the breakthrough pressures of the neighboring pores (see Equation~\ref{eqbreakthrough}). Additionally, the flow entering the membrane matrix possesses impulse in the axial direction, increasing the local kinetic pressure in axial direction.
      
        Second, there might be an axial transport of the aqueous phase inside the polymer matrix of the membrane fibers. On the one hand, this might be induced due to shear flows at the membrane lumen and shell side. On the other hand, part of the feed flow might enter the membrane at the feed side through the open polymer matrix cross-section. Although the pressure drop across the membrane in the axial direction is very high, local mixing effects and miscibility of feed solution with the pore stabilizing glycerol might lead to a more preferential wetting in the first part of the module. 
        
        As commercial membrane modules mostly possess a module length significantly higher than the module investigated here, it seems likely that this effect is even more pronounced in commercial membrane modules. In industrial applications, filtration cycles may lie in the same order of magnitude as the wetting processes observed in the scope of this study. 
        
        Hence, the findings regarding the wetting progress revealed in this study can significantly affect the membrane performance in industrial membrane processes. However, the gradient in the wetting degree over the module length needs further investigation in future studies to fully understand the axial progression of membrane wetting.
        
    \subsection{Aqueous solution wetting ethanol-washed and dried (air-filled) membranes}
        
\begin{figure}[H]
                \centering
                \includegraphics[width=\columnwidth]{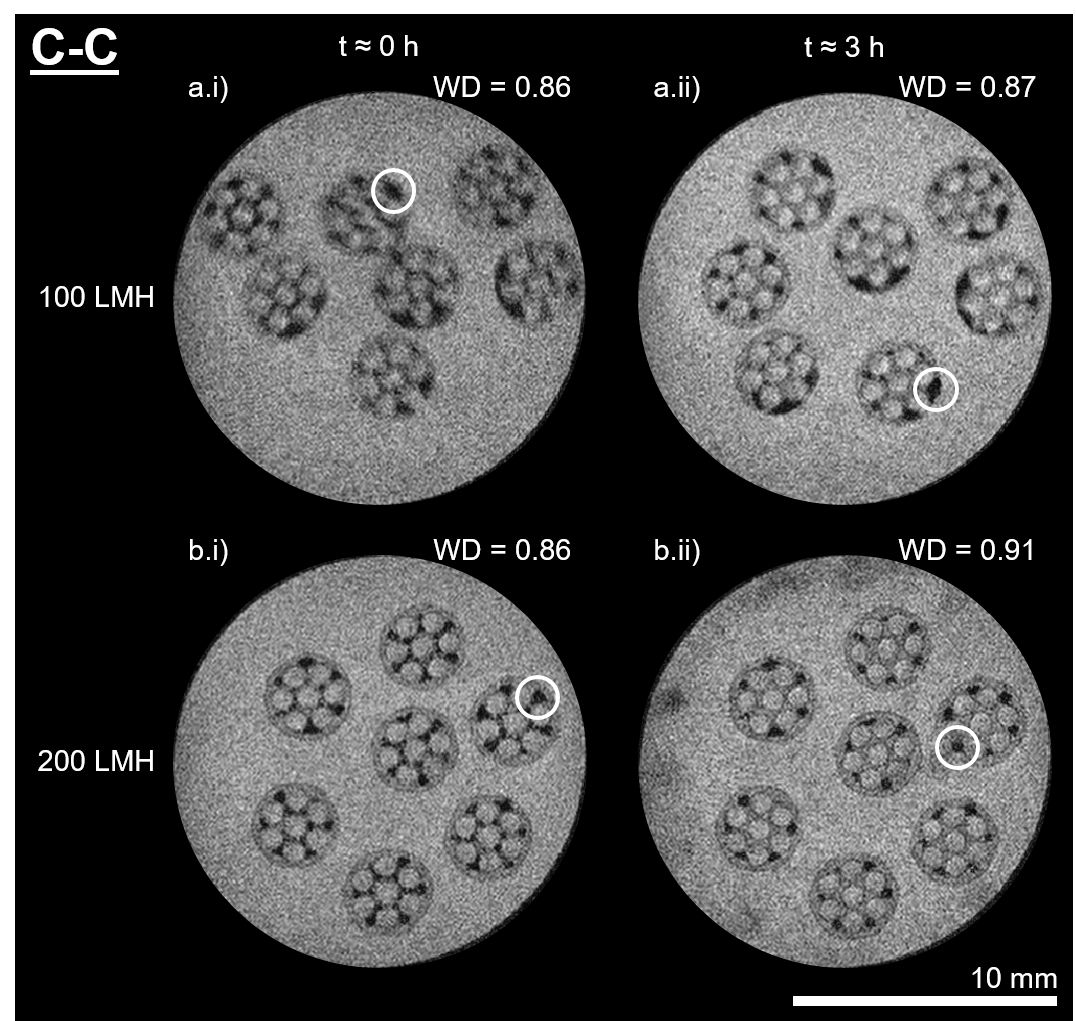}
                \caption{Magnetic resonance images of seven fiber modules containing ethanol washed and dried membranes operated at 100~LMH (top) and 200~LMH (bottom) in the initial stage of the experiment (left) and after approximately three hours (right). White-circled areas highlight the different wetting behavior compared to untreated membranes (see Figure~\ref{figFlowPaths}).}
                \label{figMRIwashed}
            \end{figure}

        The wetting behavior of the membrane pores depends on the invading fluid's properties, as well as the defending fluid that is displaced during the wetting procedure. After investigating influence parameters on the invading fluid, namely flux and packing density, we also aimed to investigate effects arising from the defending fluid properties. Therefore, in a first step, we removed the factory-filled glycerol inside the pores by washing the membranes with ethanol. As indicated in literature~\cite{Kochan.2009}, the material system is stable in ethanol. To wash the membranes, they are submerged in ethanol for 24 hours. For the washing process, approximately 150~mL of ethanol is required. Subsequently, the ethanol-soaked fibers were dried. Afterward, the air inside the polymer matrix was displaced with the same experimental parameters as described above. In Figure~\ref{figMRIwashed}, results are displayed for permeation experiments at 100~LMH (top) and 200~LMH flux (bottom) in the initial state (left), as well as after three hours of permeation (right). The temporal evolution can be seen in Supplement Figure~7.
        
        Initially, the membranes swell during the first contact with aqueous solution. This swelling causes significant bending of the fibers, which results in blurred visualization from MRI measurements, as shown in Figures~\ref{figMRIwashed}~a.i). The initial wetting pattern distinctly differs from the behavior observed with factory-filled fibers. Here, the circular region next to the outer membrane skin is immediately wetted with aqueous solution, independent of the applied flux. The rest of the polymer matrix remains non-wetted. 
        
        After three hours of permeation, non-wetted domains remain in the membrane fibers for both applied fluxes, as shown in Figures~\ref{figMRIwashed}~a.ii) and b.ii). For a flux of 100~LMH, the remaining non-wetted domains are unevenly distributed. This asymmetry is in contrast to the reference experiments with factory-filled fibers. One potential reason for these non-uniform wetting patterns is a partial collapse of the active skin pore system throughout the washing procedure. As glycerol is assumed to be factory-filled as a pore stabilizer, the drying step after glycerol removal might have led to a mechanical change of the polymer matrix. A change in some parts of the polymer matrix can lead to non-uniform transport resistances in the membrane. These non-uniformities would cause non-uniform flow distributions and, hence, non-uniform wetting patterns. A potential second cause is the more distinct membrane movement observed at the beginning of the experiment with 100~LMH flux. However, the wetting pattern observed after three hours of permeation with a flux of 200~LMH is symmetric. Like before, the outer membrane skin is entirely wetted. Similar to the experiments discussed before, this leads to fully wetted areas between the outer lumen channel and the outer membrane skin. In contrast to the reference experiments with factory-filled fibers (see section \ref{secGlycfilled}), the polymer matrix domains with larger pore sizes remain non-wetted. For reference, these specific domains, again, are highlighted by white circles in Figure~\ref{figMRIwashed}.
        
        Especially the observation of non-wetted domains with large pore sizes even for high fluxes is unexpected. With PES being a hydrophobic material, it would be more intuitive to wet larger pores more easily than narrow pores, where the necessary local pressure gradient needs to be higher. We conclude that the feed flow bypasses the polymer matrix domains with larger pore sizes that are mainly contributing to the mechanical stability of the fiber to a large extent. This way, the local fluid pressure acting on the air-filled pores does not reach the breakthrough pressure. Also, the difference in wetting patterns might be explained if there is a feed flow entering the membrane fibers in the axial direction through the open cross-section of the polymer matrix at the feed inlet. For the glycerol pre-wetted fibers, aqueous feed solution and glycerol's miscibility might lead to a steady axial wetting of the membrane fiber in the axial direction. As air macroscopically is not miscible with the feed solution, this mixing effect cannot contribute to the overall fiber wetting. However, the exact flow conditions during wetting in radial and axial direction need further investigation in future studies. Setting a corresponding CFD model would be useful to estimate temporal pressure and flow distributions, and thus, draw conclusions on the wetting behavior.
        
    \subsection{Aqueous solution wetting ethanol-filled membranes}
        In the section above, the membrane pore system might have been altered due to the drying procedure. To observe the feed wetting progression without any distortion by pore system altering, we investigated the wetting behavior with aqueous solution in ethanol-filled membranes. 
        
        An ethanol flux of 100~LMH was applied to the membrane module. This washing step represents a displacement procedure of glycerol by another fluid. Hence, we tried to investigate the wetting kinetics and wetting patterns for this displacement procedure. However, monitoring this displacement in MRI lead to results that are not suitable for image processing analysis (see Supplement Figure~3). There are three reasons for this reduction in image quality for ethanol displacing glycerol. First, the measuring time needs to be increased to approximately 40~min per image as the MRI signal of ethanol is only about a tenth of the water signal. Second, wetting kinetics are faster for ethanol displacing glycerol than for aqueous solution displacing glycerol. Hence, the temporal resolution is not sufficient for monitoring this wetting process. Third, the membranes bend significantly when flushed with ethanol due to swelling effects. This induces membrane movement during the MRI measurement, which leads to a further signal noise in the obtained data. 
 
\begin{figure}[H]
                \centering
                \includegraphics[width=\columnwidth]{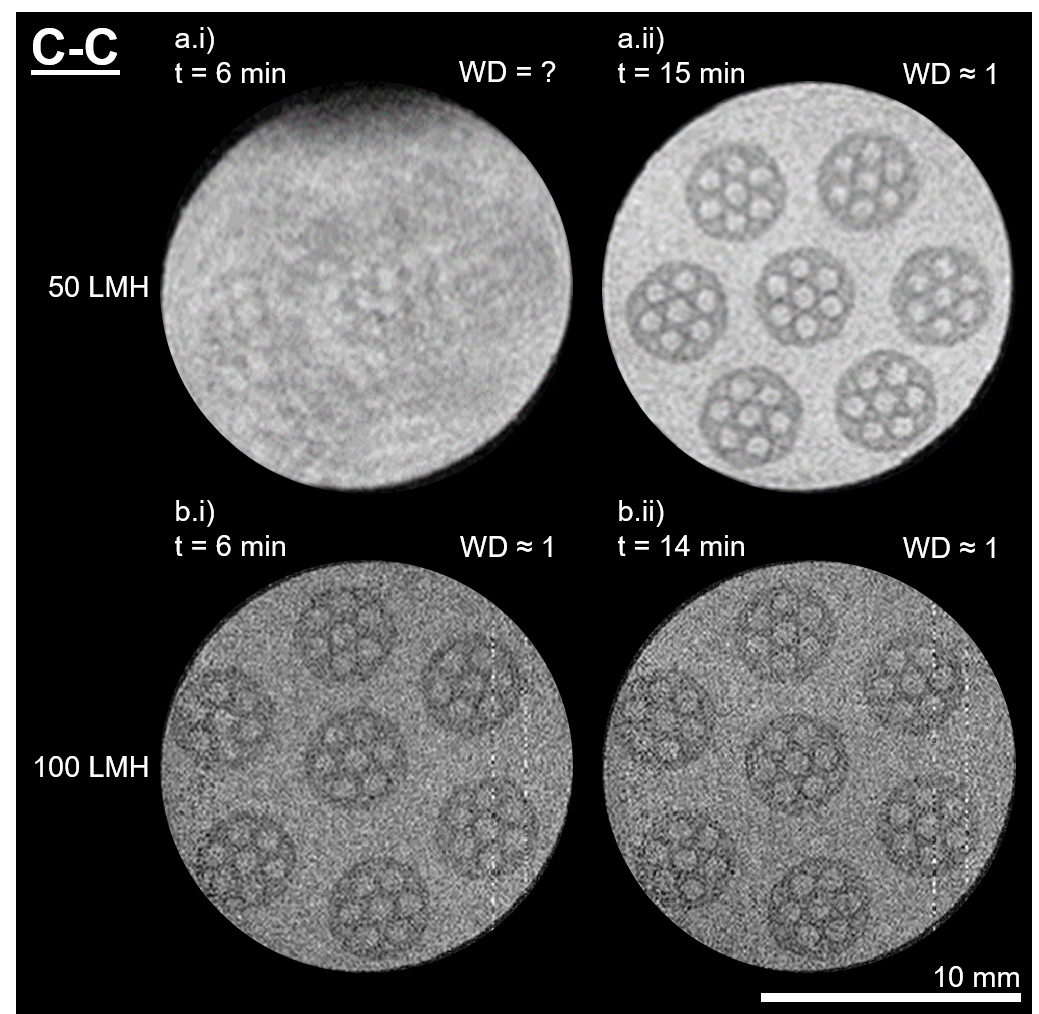}
                \caption{Magnetic resonance images of seven fiber modules containing ethanol filled membranes taken during the initial stage of the displacement process of ethanol with \ce{CuSO4}-solution at 50~LMH (top) and 100~LMH (bottom) at position C-C.} 
                \label{figMRIDisp}
            \end{figure}

        After full wetting of the membrane fibers with ethanol, aqueous solution was used for ethanol displacement with 50 and 100~LMH, respectively. In Figures~\ref{figMRIDisp}~a.i) and a.ii), results are shown for aqueous solution displacing ethanol with a reduced flux of 50~LMH. At the first contact with the aqueous solution, the membranes start to move significantly due to swelling effects. This movement causes the blurry image shown in Figure~\ref{figMRIDisp}~a.i). After approximately ten minutes, the movement of the fibers has ceased. The subsequent measurement shows an evenly strong signal throughout the whole cross-sectional area of the SevenBore\texttrademark{} fibers. With a flux of 100~LMH, even the initial MRI measurement after shell-sided module flooding revealed thoroughly wetted polymer structures. 
        
        Two reasons were identified for this tremendously faster wetting with aqueous solution compared to differently pre-treated membranes. First, ethanol has a significantly lower viscosity than glycerol, a lower surface tension, and a lower contact angle on PES surfaces~\cite{Kochan.2009}. This lower surface tension reduces the necessary breakthrough pressure for pore wetting. Second, the polymer matrix itself mechanically reacts to the immersion in ethanol by swelling. The swelling leads to more free volume inside the polymer matrix and increases the permeability for a water-rich phase by diffusion. This increase, in turn, causes faster transport of aqueous solution throughout the membrane and, hence, faster wetting kinetics. However, a remaining membrane swelling causes a transport resistance for the convective aqueous flux and might lead to lower fluxes compared to differently pre-wetted membranes. 
                
    \subsection{Temporal evolution of obtained flux, transmembrane pressure (TMP), wetting degree (WD), and membrane resistance (R)}    
\begin{figure}[H]
\centering%
\begin{minipage}{0.45\linewidth}
  \includegraphics[width=\linewidth]{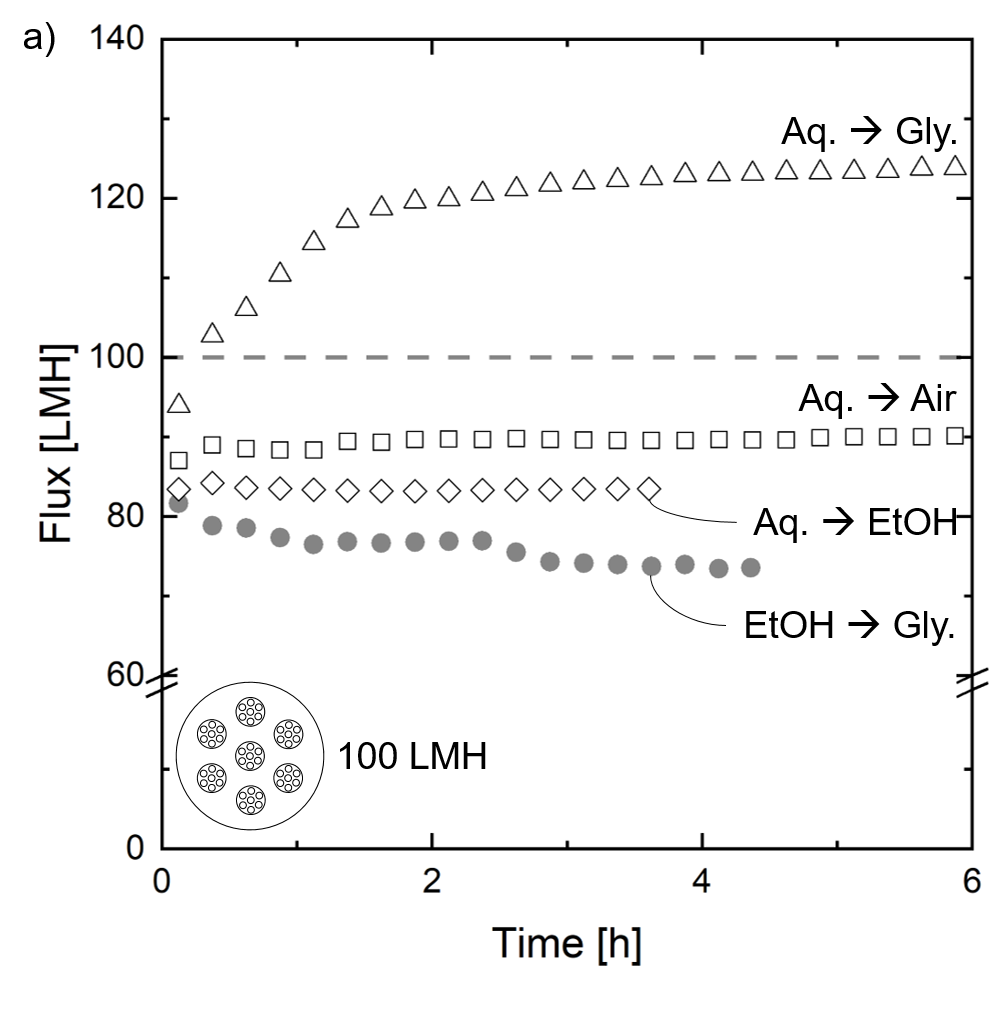}
\end{minipage}\qquad
\begin{minipage}{0.45\linewidth}
  \includegraphics[width=\linewidth]{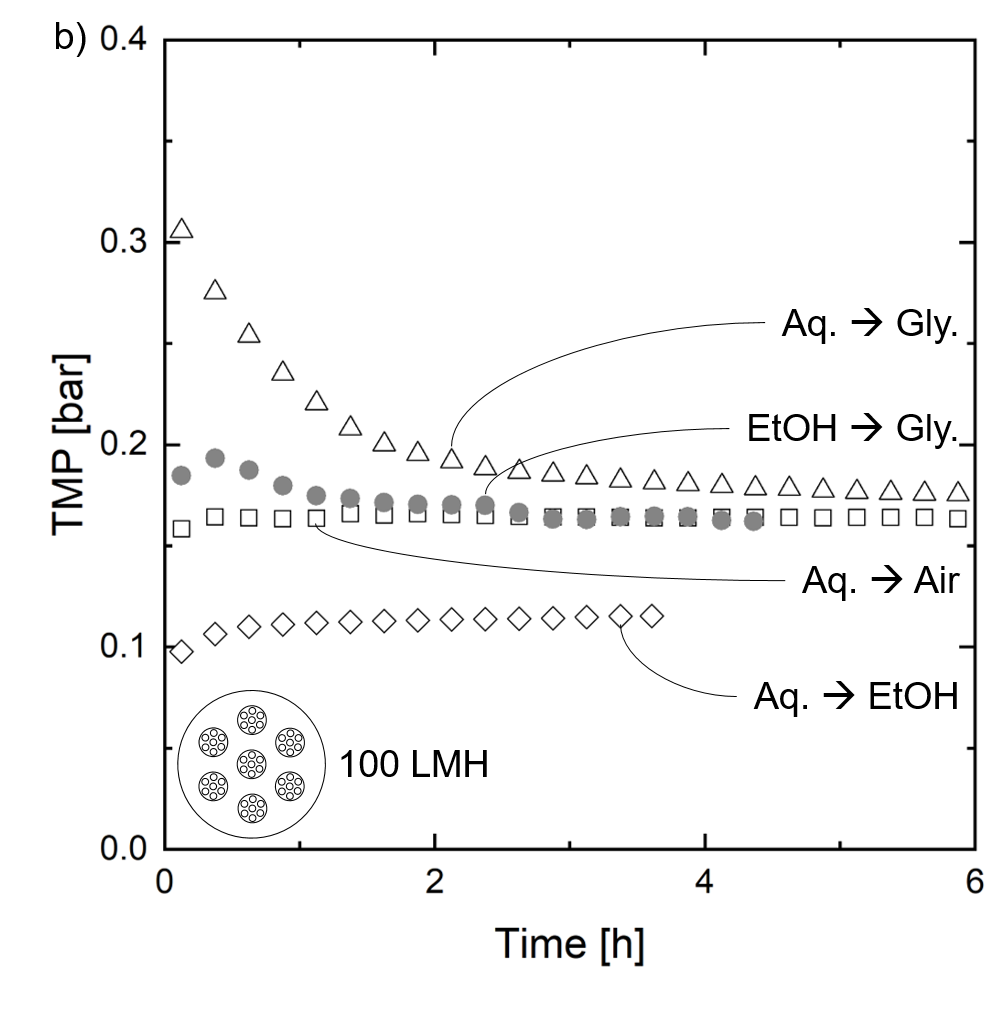}
\end{minipage}\vspace{0.2cm}
\begin{minipage}{0.45\linewidth}
  \includegraphics[width=\linewidth]{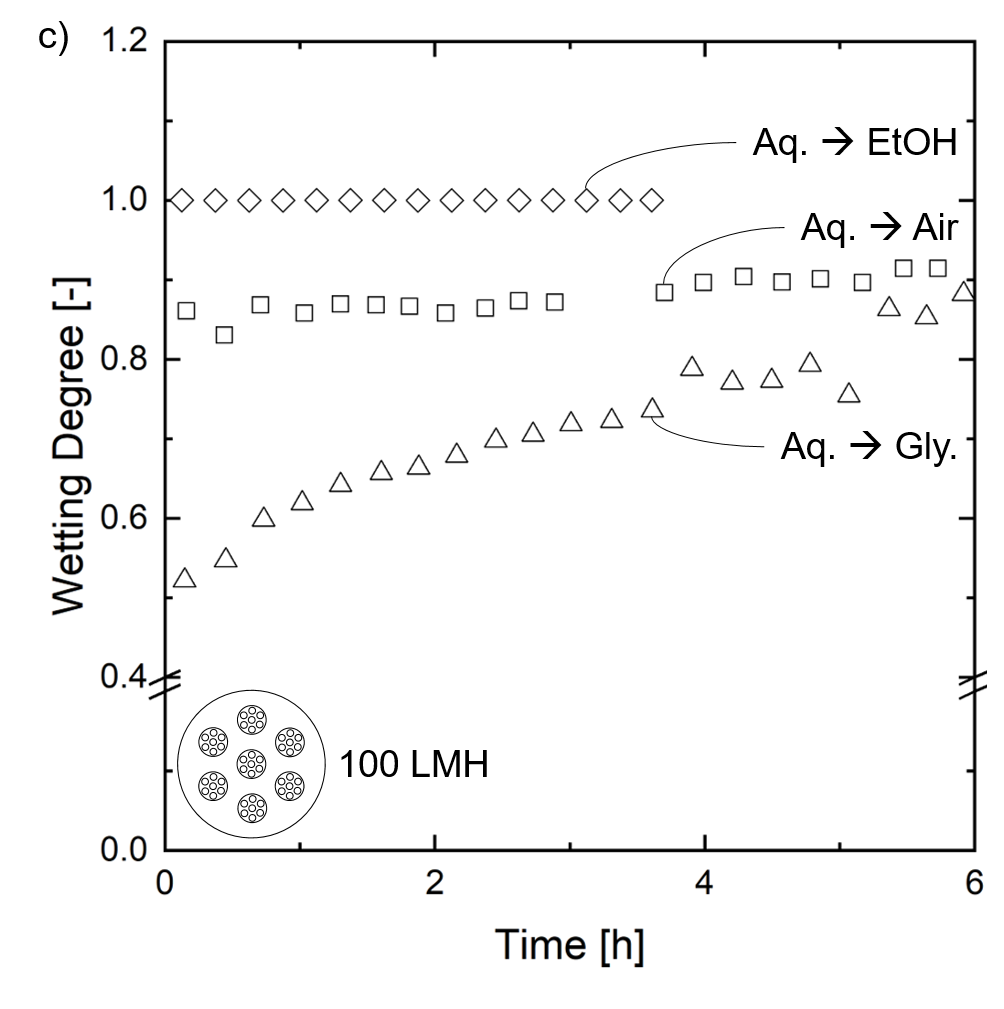}
\end{minipage}\qquad
\begin{minipage}{0.45\linewidth}
  \includegraphics[width=\linewidth]{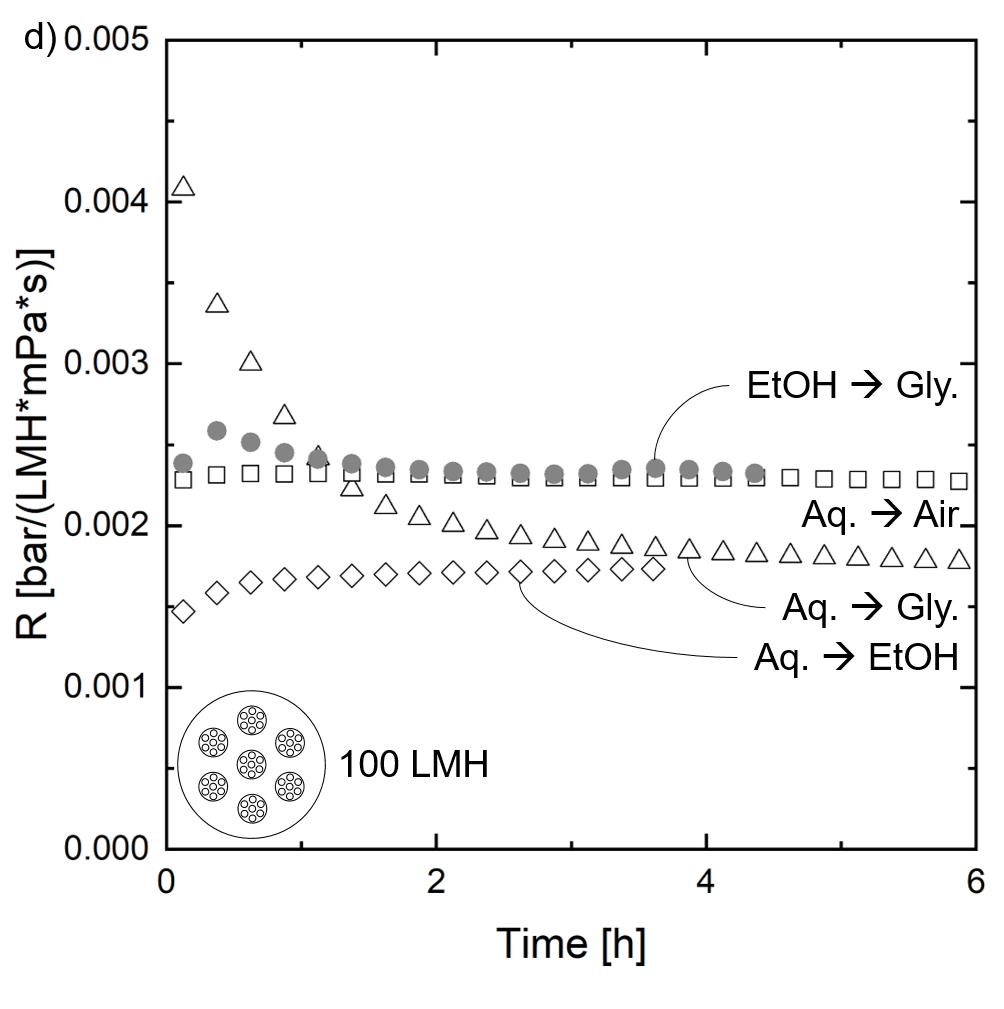}
\end{minipage}
  \renewcommand\thefigure{\arabic{figure}}
\caption{Temporal evolution of (a) the measured flux, (b) the measured correlating TMP, (c) the monitored wetting degree in C-C, and (d) the calculated membrane resistance during the wetting of seven fiber modules at 100~LMH with an initial value of 85~LMH. Investigated material systems are delivery-state membrane fibers wetted with aqueous solution~(\protect\markertriangleunfilled), washed and dried (air-filled) membrane fibers wetted with aqueous solution~(\protect\markersquareunfilled), untreated membrane fibers wetted with ethanol~(\protect\markergraycircle), and ethanol-filled membrane fibers wetted with aqueous solution~(\protect\markerdiamondunfilled).}
\stepcounter{figure}
\label{figFluxTMPComp}
\end{figure}

        Besides the phenomenological observation of local membrane wetting patterns, we also observed the macroscopic quantities flux and TMP of the permeation. Figure~\ref{figFluxTMPComp} shows the course of flux (Figure~\ref{figFluxTMPComp}~a)), TMP (Figure~\ref{figFluxTMPComp}~b)), calculated membrane resistance (Figure~\ref{figFluxTMPComp}~c)), and wetting degree obtained from MRI data (Figure~\ref{figFluxTMPComp}~d)) over time for a seven fiber module with a target flux of 100~LMH for the investigated material systems. Inside the polymer matrix, glycerol is displaced by the aqueous solution~(\markertriangleunfilled), air is displaced by the aqueous solution~(\markersquareunfilled), and ethanol displaced by the aqueous solution~(\markerdiamondunfilled). For reference, the measured data during the initial displacement of glycerol by ethanol is dispayed~(\markergraycircle). For a higher flux, the graphs' characteristics remain the same, although the absolute values change (see Supplement Figure~4 and Figure~5).

        Since a ISM405A gear pump was used for all experiments conducted in this study and we did not use any other control loop, neither an ideal constant flux nor an ideal constant pressure mode was realized. The initial flux at the beginning of the experiment was set to 85~LMH. During the experiments, wetting of the membranes led to changes in flux as well as in TMP.  However, the behavior of flux and TMP over time can still be used for qualitative wetting characterization. From this data, we calculated the membrane resistance R as described in Equation~\ref{eqResistance} in the function of the applied TMP, the measured flux J, and the viscosity of the invading fluid $\eta$.
        
        \begin{equation}
        \label{eqResistance}
           R=\frac{\text{TMP}}{J*\eta}
        \end{equation}        
        
        Figure~\ref{figFluxTMPComp}~a) shows the flux over time for the three investigated material systems with a target flux of 100~LMH. Also, the progression of flux during the displacement of glycerol by ethanol is displayed for reference. The flux remains mostly constant over time for aqueous solution perfusing washed and dried membranes~(\markersquareunfilled, air-filled) and ethanol-filled membranes~(\markerdiamondunfilled). In contrast, there is a significant increase in flux for the factory-filled membranes permeated with the aqueous solution~(\markertriangleunfilled). In general, this is in good agreement with the optical analysis discussed above since the air-filled, and the ethanol-filled membranes show mostly wetted membrane skins from the beginning of the experiments. Hence, there is no increase in the active membrane area over time, resulting in stable membrane performance. During the permeation with ethanol~(\protect\markergraycircle), the flux constantly decreases, which can be accounted for polymer swelling.
        
        Concerning the course of the applied TMP displayed in Figure~\ref{figFluxTMPComp}~b) for the ethanol permeation (\markergraycircle), only the drop in flux between two and three hours of permeation time can be reasoned. At the beginning of the experiment, the TMP slightly increases before returning to the initial value, as does the membrane resistance. We conclude that flux and TMP courses are mainly dominated by swelling and altering phenomena of the polymer matrix immersed in ethanol. After four hours of permeation, a steady state is reached in all experiments displayed in Figure~\ref{figFluxTMPComp}. Here, the permeation with ethanol through factory-filled membranes~(\markergraycircle) shows the lowest flux at a similar applied TMP. The transport resistance is highest as displayed in Figure~\ref{figFluxTMPComp}~d), which is caused by an altered pore size distribution caused by swelling effects. In contrast, no change in flux and TMP over time was observed for the aqueous solution displacing ethanol~(\markerdiamondunfilled). In combination with the MRI results presented in Figure~\ref{figFluxTMPComp}~c), this hints to a fast and complete wetting progression due to good miscibility and low viscosity.
        
        No significant change in either flux or TMP was observed for the ethanol-washed, dried, and subsequently perfused membranes (\markersquareunfilled, air-filled membranes). Although the optical evaluation via MRI revealed large parts of the membrane domains with larger pore sizes as non-wetted (see Figure~\ref{figMRIwashed}), these appear to be no limiting factor for the mass transport in the absence of a particular matter. However, drying the membrane without a pore stabilizing agent might have altered the active membrane skin morphology. Potentially, the pore network system in the active layer has collapsed, leading to larger pore diameters. This collapse would reduce the necessary breakthrough pressure (see Equation~\ref{eqbreakthrough}) and explains the low membrane resistance (see Figure~\ref{figFluxTMPComp}~d)) at the beginning of the permeation experiments compared to other fluid systems. Towards the end of the permeation experiments, the resistance of the air-filled fiber~(\markersquareunfilled) remains the same, whereas the resistance of the factory-filled fiber~(\markertriangleunfilled) has dropped. The higher resistance level might be caused by the non-wetted areas that locally require higher flow rates inducing an additional transport resistance. Also, the reduced flux of the air-filled compared to factory-filled membranes at steady-state of permeation might be reasoned with a reduced number of pores in a collapsed pore network system. However, this reasoning requires spatial resolution of flow patterns and in-depth active layer investigation, which need to be addressed in future studies.
        
        The increase in flux for the factory-filled membranes~(\markertriangleunfilled) at the beginning of permeation with aqueous solution complies with a decrease in TMP. As well as in the evaluation of the wetting degree over time (see Figure~\ref{figFluxTMPComp}~c)), a flattening of the curve displayed in Figure~\ref{figFluxTMPComp}~a) can be observed after approximately two hours. At this point in the permeation experiment, the optical evaluation via MRI shows the first macroscopic connection of wetted domains between the lumen channels and the shell side of the module (see Supplement Figure~6, wetting degree as a function of time and applied flux for delivery-state membrane fibers wetted with aqueous solution). From this point onward, the further increase in flux can be reasoned with a broadening of the existing wetted fluid flow pathways in radial and axial directions.
        
\section{Conclusion}

    We use MRI as an in-situ characterization of macroscopic wetting patterns inside polymeric hollow fiber SevenBore\texttrademark{} membranes. The effects of packing density, applied flux, lateral position, and membrane pretreatment are elaborated. For quantitative analysis, an image processing tool is developed to calculate the membranes' wetting degree. This study helps to understand (a) complex wetting phenomena inside multibore membranes in dead-end permeation, (b) the membranes' interaction with their surroundings due to neighboring membranes, and (c) the effect of the used fluid system for displacement on the resulting wetting patterns.
    
    We show that the wetting procedure of these commercial membranes takes several hours when conventionally flushed with aqueous solution, even with high fluxes of 200~LMH. The membrane performance is tremendously decreased, as not all theoretically possible active area is available for mass transport for the duration of this displacement procedure. 
    
    In general terms, the wetting progresses inside-out while building wetted short-cut pathways between the outer lumen channels and the membranes' outer skin. Wetting patterns change drastically by changing the flux from 100~LMH to 200~LMH, and the wetting kinetics increase with increasing packing density.This increase in wetting kinetics can be reasoned with higher flow rates at the shell side, inducing different pressure gradients across the membrane. 
    
    In addition to a successive wetting in the radial direction, we revealed an axial wetting direction by comparing the wetting patterns at different lateral positions inside the module. This axial wetting, again, can be reasoned with changing shell-sided flow conditions along the module length. This phenomenon needs further in-depth investigation and will be subject to future studies.
    
    To compare wetting patterns and kinetics, we investigated the wetting behavior of ethanol-washed, dried, and subsequently filtrated membranes. The arising wetting patterns differ in two striking ways. First, the resulting wetting patterns have a less symmetric shape. Second, the membranes' outer area was immediately wetted, while some inner parts of the membrane remained air-filled. In contrast, the complete membrane cross-section is almost immediately wetted when displacing ethanol inside the polymer matrix by aqueous solution. However, the initial immersion in ethanol causes significant polymer swelling that displays an additional transport resistance. 
    
    This study reveals the importance of flow conditions during the first cycle of membrane filtration on the membrane wetting efficiency. The resilient patterns of non-wetted membrane domains hint towards an optimization potential for current industrial modes of operation. Hence, this work pioneers finding strategies for membrane performance optimization by more active wetting control. 
    
 \section*{Acknowledgment}
    M.W. acknowledges DFG funding through the Gottfried Wilhelm Leibniz Award 2019 (grant ID = WE 4678/12-1). M.W. acknowledges the support through an Alexander-von-Humboldt Professorship and the European Research Council (ERC) under the European Union's Horizon 2020 research and innovation program (grant agreement no. 694946). This work was also performed in part at the Center for Chemical Polymer Technology CPT, which is supported by the EU and the federal state of North Rhine-Westphalia (grant no. EFRE 30 00 883 02). The authors thank Geert-Henk Koops (Suez) for providing membrane materials. The authors thank Lorenz Aldefeld and Hannah Buchholz for their constructive effort in this work and Karin Faensen for her support in electron microscopy.

\bibliographystyle{elsarticle-num}
\biboptions{sort&compress}
\bibliography{Publication1}{}

\end{document}